# Density-functional study of electronic structure of quantum dots in high magnetic fields

A.A. Vasilchenko

*National Research Tomsk State University, 634050 Tomsk, Russia*

Abstract

Density-functional theory is used to study the electronic structure of quantum dots in a magnetic field. New series of magic numbers are found for the total angular momentum of electrons. The empirical formula for the plateau width is obtained. The effect of a charged impurity on the electronic structure of a quantum dot has been studied.

## 1. Introduction

The most impressive results on taking into account the electron-electron interaction in two-dimensional quantum dots were obtained in [1–9], in which the many-particle Schrödinger equation was solved numerically. Results of numerical solution of the many-particle Schrödinger equation for the number of electrons $N <$ 10 showed that the energy spectrum of electrons has interesting features. In particular, the ground and metastable states of a multielectron system in a magnetic field are observed only at certain values of the total angular momentum of electrons $M$. These values of total angular momentum are known as magic numbers. The authors of [1] found that magic numbers have period $\Delta M = N$. Later, a new series of magic numbers with period $\Delta M = N-1$ was found [5, 9]. If the ground state of electronic system corresponds to magic numbers with a period $\Delta M = N-1$, then electrons have a polygonal electron configuration in which one electron is in the center. The results of calculations [1 - 9] showed that energy spectrum of electrons in quantum dots has a gap, whose origin is associated with electron-electron interactions. Exact calculations have shown that, at least at the filling factor of Landau level $\nu < 0.4$, electrons crystallize and form a Wigner molecule [5].

Numerical diagonalization of the many-particle Hamiltonian requires high computational costs, and calculations can only be carried out for no more than ten electrons. At present, one of the most powerful methods to take into account the many-particle interaction is the density-functional theory (DFT). In this work, we study the electronic properties of two-dimensional quantum dots in a perpendicular magnetic field using DFT.

## 2. Theoretical model

We use the effective system of atomic units in which the energy is expressed in units of Ry = $e^2/(2\varepsilon a_B)$, and the length in units of $a_B = \varepsilon\hbar^2/(m_e e^2)$, where $m_e$ is the effective electron mass, $\varepsilon$ is the dielectric constant. All calculations are performed for two-dimensional quantum dots based on GaAs, for which $\varepsilon$ = 12.4 and $m_e$ =0.067$m_0$ ($m_0$ is a free electron mass). For GaAs, we get $a_B$ = 9.8 nm, $Ry$ = 5.9 meV.

According to DFT, the total energy of the electron system in an external potential $V_{ext}(r)$ is a single-valued electron density functional $n(r)$:

$$E[n] = T[n] + E_{ext}[n] + E_H[n] + E_{xc}[n], \qquad (1)$$

where $T[n]$ is kinetic energy of non-interacting electrons in a magnetic field $B$, which is given by vector potential $A=B(-y/2, x/2, 0)$.

The second term in expression (1) is related to the external interaction and in the two-dimensional case is given by

$$E_{ext}[n] = \int V_{ext}(r) n(r) dr, \qquad (2)$$

where $V_{ext}(r) = \dfrac{\omega_0^2 r^2}{4} - \dfrac{2z_0}{r}, \qquad (3)$

The external potential $V_{ext}(r)$ is created by an impurity with charge $z_0$ and a parabolic potential.

The Coulomb energy has the following form

$$E_H[n] = \frac{1}{2}\int V_H(r) n(r) dr, \qquad (4)$$

where $V_H(r) = 2\int_0^\infty \dfrac{n(r')}{|r-r'|} dr', \qquad (5)$

The form of the exchange-correlation energy $E_{xc}[n]$ is unknown. Further, we exclude the self-interaction of electrons and take into account only the exchange energy, for which we use the local-density approximation:

$$E_x[n] = \int \varepsilon_x(n) d\mathbf{r} - \sum_m \int \left( \varepsilon_x(n_m) + \frac{1}{2} V_{H,m}(r) \right) n_m(r) d\mathbf{r}, \qquad (6)$$

where $V_{H,m}(r) = 2\int_0^\infty \frac{n_m(r')}{|\mathbf{r}-\mathbf{r}'|} d\mathbf{r}'$, $n_m(r)$ is $m$-th electron density, $\varepsilon_x(n)$ is exchange energy per electron for a homogeneous electron gas, which for the lower Landau spin level has the following form:

$$\varepsilon_x(n) = -\sqrt{2\pi}\pi L n(r), \qquad (7)$$

where $L$ is magnetic length.

Calculations are carried out for magnetic fields in which two-dimensional electrons are spin-polarized. By varying the energy (1) and taking into account the circular symmetry, we obtain the Kohn-Sham equations for spin-polarized electrons:

$$\left\{ -\frac{\partial^2}{\partial r^2} - \frac{1}{r}\frac{\partial}{\partial r} + \frac{r^2}{4L^4} + \frac{m^2}{r^2} - \frac{m}{L^2} + V_{eff}(r) \right\} \psi_m(r) = E_m \psi_m(r), \qquad (8)$$

with an effective one-particle potential

$$V_{eff}(r) = V_H(r) - V_{H,m}(r) + 2\alpha(n(r) - n_m(r)) + V_{ext}(r), \qquad (9)$$

where $m$ is the angular momentum of an electron, $n_m(r) = |\psi_m(r)|^2$, $n(r) = \sum_{occ\ m} n_m(r)$, $\alpha = -\sqrt{2\pi}\pi L$.

## 3. Numerical results and discussion

The nonlinear system of Kohn-Sham equations (8-9) is solved numerically. In Figure 1, calculation results are compared with the results of exact diagonalization [5]. It can be seen that energy value calculated using DFT is approximately 13% higher than the exact value, and positions of the energy minima coincide. Most importantly, the same magic numbers are obtained and the period $\Delta M = N - 1$, as in exact calculations. In the one-particle approximation, this means that one electron is located at the center of the quantum dot, while the rest electrons are distributed over the ring. For example, for a state at $M = 99$, electron configuration is (0, 14, 15, 16, 17, 18, 19).

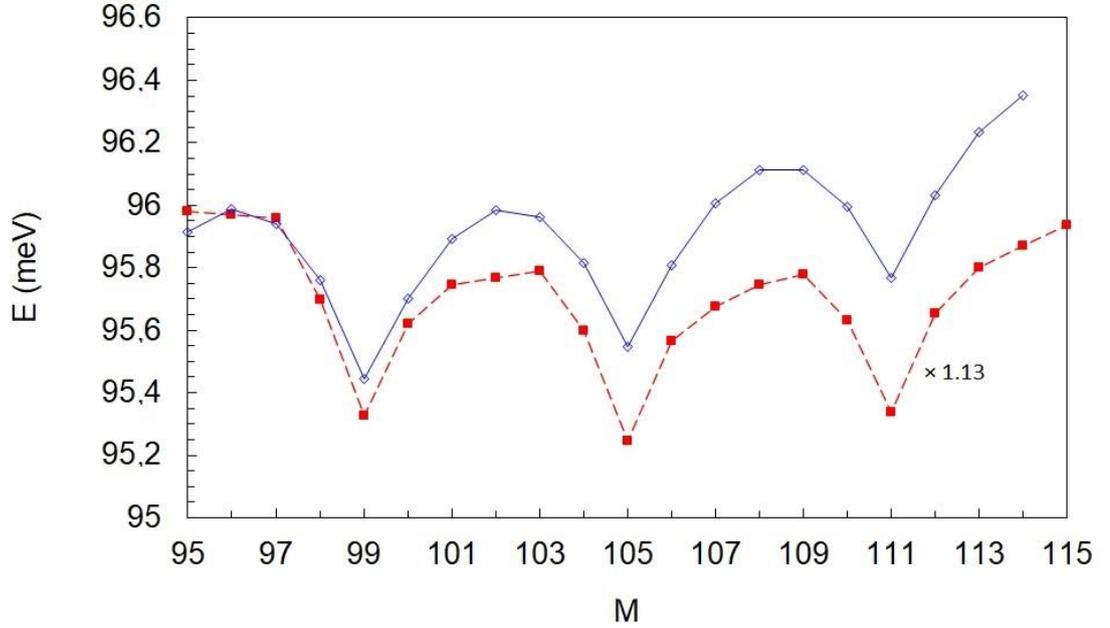

Figure 1. Dependence of energy on the total angular momentum of electrons; $N=7$, $B =18.8$ T, $\hbar\omega_0 =3$ meV. The zero of energy corresponds to $N(\omega_0^2+1/L^4)^{1/2}$. ■ – exact result [5], ○ – DFT. The lines are to guide the eye.

Figure 2 shows the dependence of the total angular momentum of electrons on the magnetic field $B$ in a quantum dot at $N = 7$ и $\hbar\omega_0 = 2$ meV. The dependence $M(B)$ has the form of a plateaus, separated by a height equal to $N$. At low magnetic fields, electrons have a compact configuration with a minimum of angular momentum $M_0 = N(N–1)/2 = 21$. The transition to the unpolarized state has not been studied, so Figure 2 shows only a part of the plateau at $M = 21$ to indicate the transition to the state with $M = 28$. With an increase in the magnetic field (up to $B = 7.3$), the compact configuration of electrons is retained and there is plateaus

$$M = M_0 + pN, \qquad (11)$$

where $p =1, 2, 3, 4, 5$.

Plateau widths at $M = 28, 35, 42$ and $49$ are close to each other. The midpoints of these plateaus $B_p$ are close to a direct-proportional dependence on the total angular momentum (straight line in Figure 2):

$$B_p = cM_p, \qquad (12)$$

where $c$ is constant.

From expressions (11) and (12), we get the following expression for the plateau widths:

$$\Delta B = \frac{2B_1}{N+1} \qquad (13)$$

where $B_1$ corresponds to the midpoint of the first plateau with $M = N(N-1)/2 + N$. Assuming that $B_0 = cM_0$ expression (13) can be written as:

$$\Delta B = \frac{2B_0}{N-1}, \qquad (14)$$

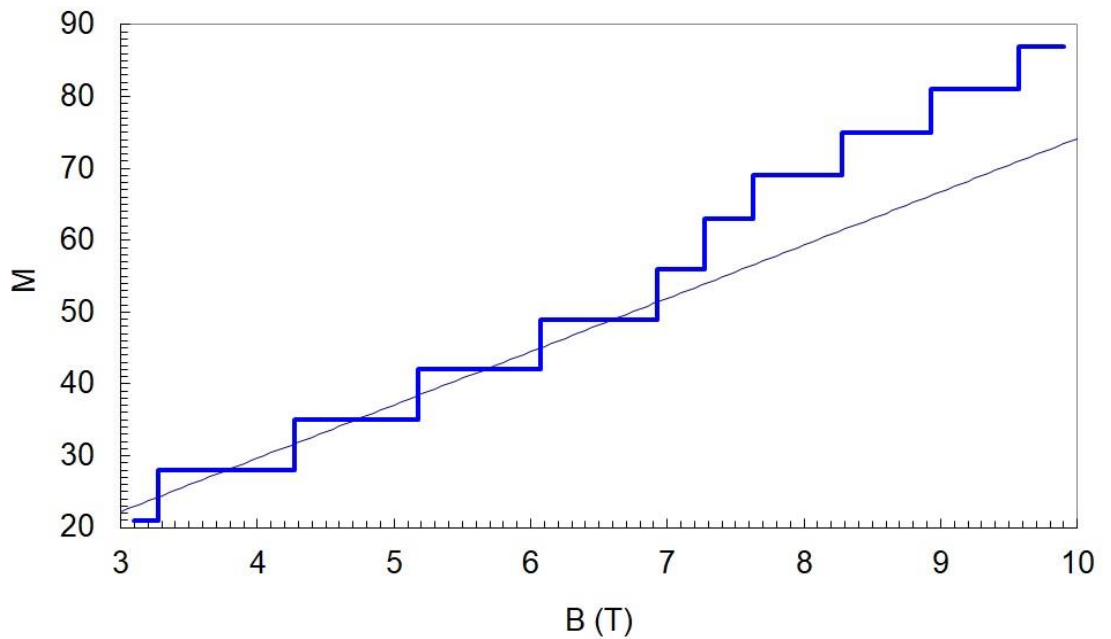

Figure 2. The dependence of the total angular momentum of electrons on the magnetic field; $N=7$, $\hbar\omega_0 = 2$ meV. The straight line corresponds to the formula (12).

The plateau with $M = 56$ has a much smaller width, and the transition to the state with $M = 63$, for which the electron configuration has the form (0, 8, 9, 10, 11, 12, 13), begins behind it. With this type of configuration, one electron is located at the center of the quantum dot, and its wave function practically does not overlap with the wave functions of other electrons. At high magnetic fields, the series of magic numbers has the form $M = M_0 + p(N-k)$ с $k = 1$ and plateau widths decrease. Note that the midpoints of these plateaus do not have a directly proportional dependence on the total angular momentum. The transition to a new series of magic numbers can be registered experimentally by studying persistent currents in quantum dots. The period of persistent current oscillations is determined from expressions (13) or (14) at low magnetic fields, and the period decreases when moving to a new series of magic numbers.

For a quantum dot with $N = 10$, a new series of magic numbers with $k = 2$ appears (Figure 3). In this case, the states with angular momenta $m = 0$ and $m = 1$ are always occupied. The sets of magic numbers (55, 65, 75), (63, 72) and (61, 69) correspond to values $k = 0, 1, 2$, respectively (Figure 3). With the increase in the magnetic field at $B = 4.6$ T, the ground state becomes the state with $M = 72$, for which the electron configuration has the form (0, 4, 5, 6, 7, 8, 9, 10, 11, 12). If $B = 5.2$ T, then there is a transition to a new series of magic numbers with $M = 85$ and electron configuration is (0, 1, 7, 8, 9, 10, 11, 12, 13, 14).

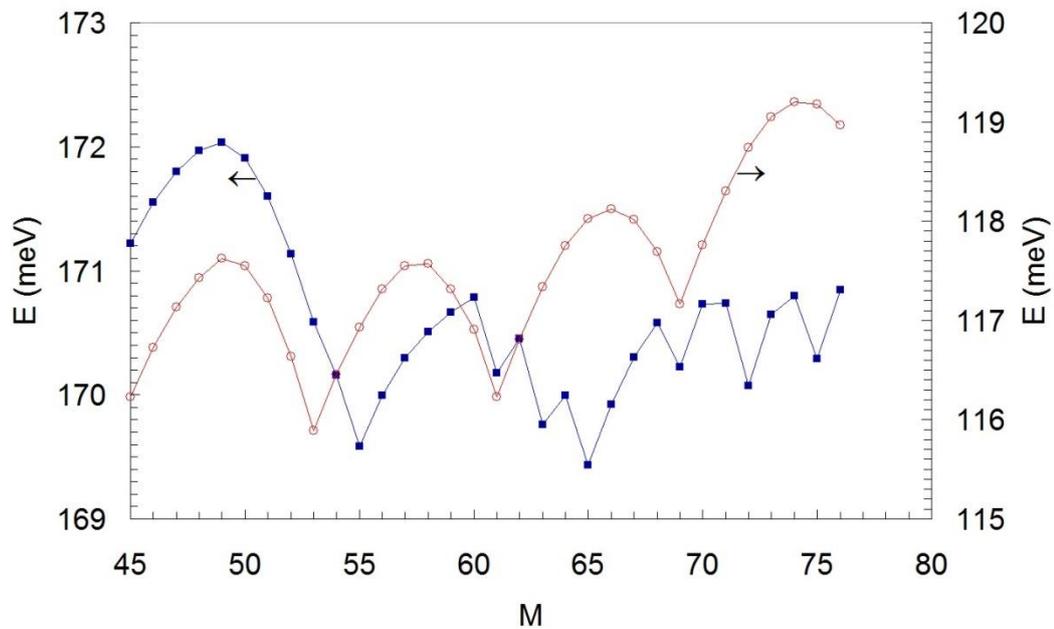

Figure 3. Dependence of the total energy of electrons on the total angular momentum of electrons; $N = 10$, $B = 4$ T, $\hbar\omega_0 = 2$ meV. ■ – $z_0 = 0$; ○ – $z_0 = 1$. The lines are to guide the eye.

We study the effect of a charged impurity on the electronic structure of a quantum dot and found that the impurity changes the ground state and sets of magic numbers for the total angular momentum of electrons. Figuries 3 shows that in the presence of a positively charged impurity, the ground state shifts towards a smaller angular momentum and the period $\Delta M = N - 2$, i.e. two electrons are localized on the impurity. In the presence of a negatively charged impurity, the ground state shifts towards a larger $M$ and only the period $\Delta M = N$ exists, so all electrons have a compact configuration.

As the number of electrons increases, new series of magic numbers appear with $k = 3, 4, 5$, due to the fact that $k$ electrons have a compact configuration and are located at the center of quantum dot, and the remaining electrons also have a compact configuration but they are distributed over the ring. Figure 4 shows the results of calculations for a quantum dot with $N = 14$. The new series of magic numbers appear, specifically: (103, 115, 127, 139) (corresponding to the period $\Delta M = N - 2$), (113, 124, 135) ($\Delta M = N - 3$), (111, 121, 131) ($\Delta M = N - 4$) and (109) ($\Delta M = N - 5$). As the magnetic field changes, not all of these states become ground states. The ground state is observed at $M = 124$ (in the case of $z_0 = 0$ for parameters shown in Figure 4). In this case, three electrons are located at the center of quantum dot, and the rest electrons are distributed over the ring. As a result of the calculations, it was concluded that the states with $k = 3$ are ground states in the range of magnetic fields from 3.5 T to 7 T (Figure 5). As the magnetic field decreases, the state with $M = 103$ with two electrons in the center of the quantum dot will become the ground state. Thus, as the magnetic field changes, not all metastable states become ground states. Metastable states affect the value of the energy gap (energy difference between the state closest to the ground state and the ground state itself) and reduce its value. For example, for $M = 124$ (Figure 4) the energy gap is 0.11 meV, and energy difference between states $M = 135$ (as the magnetic field increases this state will be the ground state) and $M = 124$ is twice as much.

Figures 4 and 5 also shows the effect of the positively charged impurity on the electronic structure of a quantum dot with $N = 14$. It can be seen that the impurity changes the magic number sets: the local minima at $M = 105, 109, 121, 131$ disappear, and the ground state moves towards a smaller total angular momentum. There are two electrons at the center of a quantum dot with an impurity at low magnetic fields. As the magnetic field increases, there are three electrons in the center of the quantum dot, the number of electrons increases to four starting from $B = 6.4$ T.

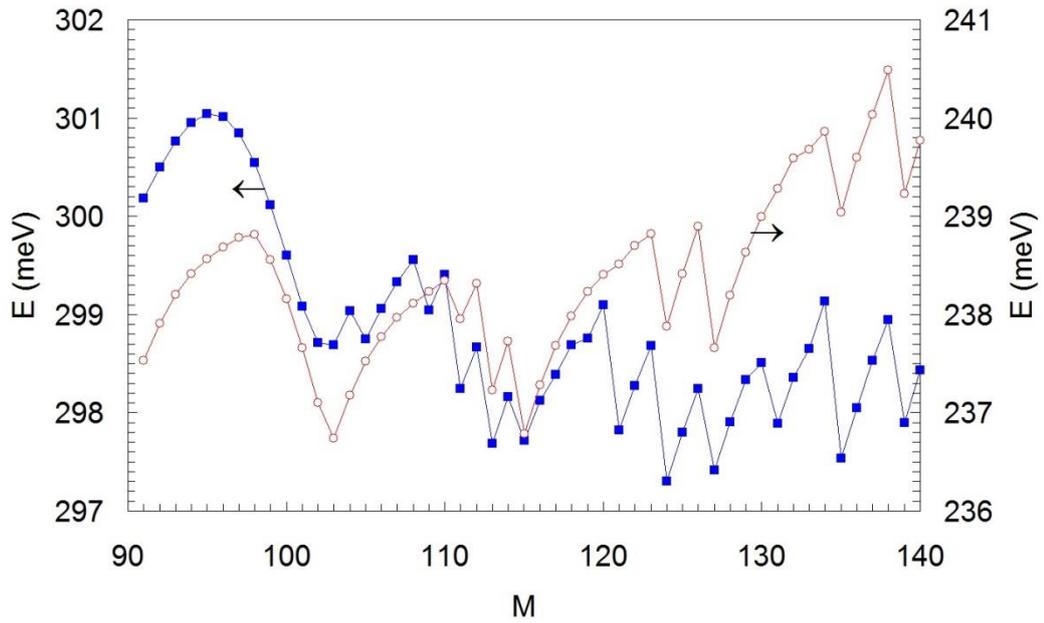

Figure 4. Dependence of the total energy of electrons on the total angular momentum of electrons; $N = 14$, $B = 4$ T, $\hbar\omega_0 = 2$ meV. ■ — $z_0 = 0$; ○ — $z_0 = 1$. The lines are to guide the eye.

It is expected that new series of magic numbers will appear and a new ring will form if the number of electrons increases. Such electron gas splitting is due to the fact that in a compact configuration of electrons, the exchange interaction reduces the energy of electrons, while the Coulomb interaction increases the energy. Note that for a compact configuration of electrons, neighboring states are occupied, so we can assume that the filling factor values will be as follows: $\nu = 1$ in the center of the quantum dot and in the ring, and $\nu = 0$ in other regions. It is assumed that for a macroscopic system, the electron gas will split into regions with

ν = 1 and ν = 0. The formation of electron drops with $N$ electrons is also possible. The filling factor can be defined as [10]

$$\nu = \frac{M_0}{M} \qquad (14)$$

In the case $M = M_0 + pN$, filling factor will always have odd denominator at even number of electrons in a drop, and in this case Bose condensation of electrons is possible. Note that in the possibility of formation of spin droplets with a number of electrons $N = 4$ and total spin $S = 2$ in zero magnetic field was experimentally shown [11].

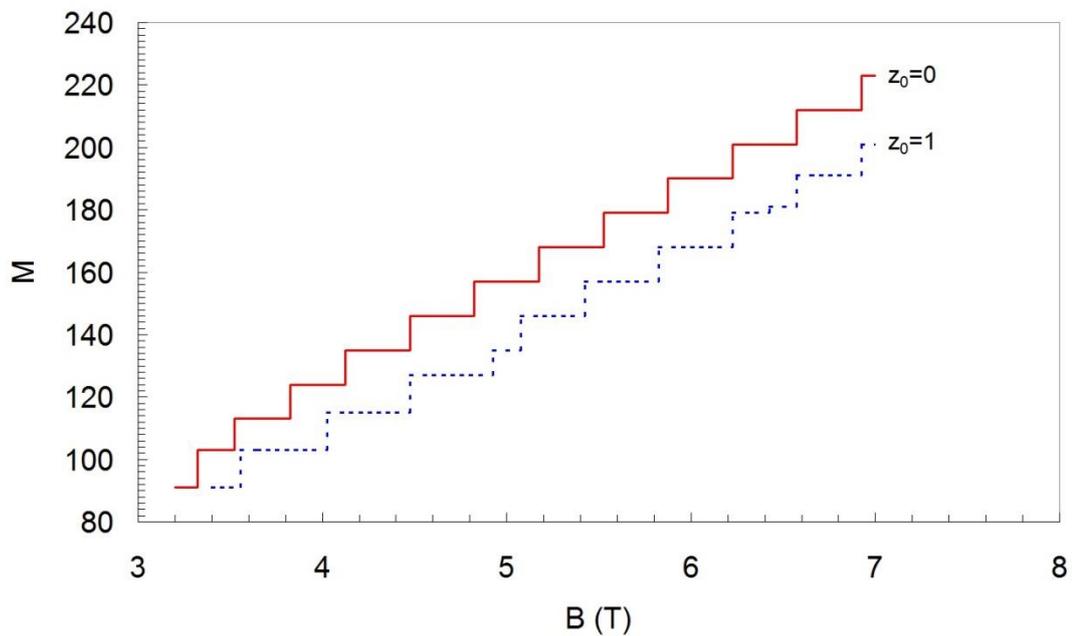

Figure 5. Dependence of the total angular momentum of electrons on the magnetic field; $N=14$, $\hbar\omega_0 = 2$ meV.

## 4. Conclusions

Theoretical study of the electronic structure of quantum dots in magnetic field was carried out using density-functional theory. A good agreement between the DFT results and the exact results for $N = 7$ has been obtained.

The new series of magic numbers with period $\Delta M = N - k$ with $k = 2, 3, 4, 5$ are found. The series of magic numbers are related to the fact that $k$ electrons have a compact configuration and they are located at the center of the quantum dot, while other electrons also have a compact configuration and are distributed over the ring. Metastable states, due to which the value of the energy gap decreases, appear with an increase in the number of electrons.

The influence of a charged impurity on the electronic structure of quantum dots has been studied, and it has been found that the impurity changes the ground state and partially destroys the set of magic numbers for the total angular momentum of electrons.

**Acknowledgements**

This work was supported by the State Assignment of the Ministry of Science and Higher Education of the Russian Federation (project No. 0721-2020-0048).

E-mail: a_vas2002@mail.ru